\documentclass[aps,prl,twocolumn,superscriptaddress]{revtex4-1}

\usepackage{graphicx}
\usepackage{dcolumn}
\usepackage{bm}

\begin{document}


\title{Tunable Surface Conductivity in Bi$_2$Se$_3$ Revealed in Diffusive Electron Transport}
\author{J. Chen,$^1$ X.~Y. He,$^1$ K.~H. Wu,$^1$ Z.~Q. Ji,$^1$ L. Lu}
\affiliation{Institute of Physics, Chinese Academy of Sciences, Beijing 100190, China}
\author{J.~R. Shi}
\affiliation{International Center for Quantum Materials, Peking University, Beijing 100871, China}
\author{J.~H. Smet}
\affiliation{Max Planck Institute for Solid State Research, D-70569, Stuttgart, Germany}
\author{Y.~Q. Li}
\affiliation{Institute of Physics, Chinese Academy of Sciences, Beijing 100190, China}


\date{\today}

\begin{abstract}
We demonstrate that the weak antilocalization effect can serve as a convenient method for detecting decoupled surface transport in topological insulator thin films. In the regime where a bulk Fermi surface coexists with the surface states, the low field magnetoconductivity is described well by the Hikami-Larkin-Nagaoka equation for single component transport of non-interacting electrons. When the electron density is lowered, the magnetotransport behavior deviates from the single component description and strong evidence is found for independent conducting channels at the bottom and top surfaces.
Magnetic-field-dependent part of corrections to conductivity due to electron-electron interactions is shown to be negligible for the fields relevant to weak antilocalization.


\end{abstract}

\maketitle

The surface of a 3D topological insulator (TI)~\cite{TIreview,Hsieh08} hosts a 2D system of Dirac electrons with spins transversely locked to their translational momenta. Such spin-helical surface states~\cite{Hsieh09a} offer a new route for realizing exotic entities such as Majorana fermions and magnetic monopoles~\cite{FuQi}. The unique surface spin structure also has profound impact on the transport properties of TI~\cite{TIreview,Ostrovsky10}. The Berry phase associated with the surface electrons causes suppression of backscattering~\cite{SuppressBS} and hence immunity to localization regardless of the strength of disorder~\cite{Ostrovsky10}. This weak antilocalization effect can be brought out by applying a perpendicular magnetic field. It produces a negative magnetoconductivity due to the breaking of time-reversal symmetry. The negative magnetoconductivity has indeed been observed in various TI thin films by several groups~\cite{ChenJ10,Checkelsky10,HeT10,LiuM10,WangJ10}. However, most of these measurements were carried out with samples in which the Fermi energy is not located in the band gap, so that they are not in the so-called topological transport regime~\cite{Hsieh09b,ChenYL09}. Since topologically trivial 2D electron systems (e.g.\ Au thin films) may also exhibit similar magnetoconductivity behavior as long as the  spin-orbit coupling (SOC) is sufficiently strong~\cite{Bergmann84}, concern has to be raised whether weak antilocalization can provide a reliable method for identifying the surface state transport, which is a key starting point for future exploration of various topological effects and novel devices~\cite{TIreview,Fu09,Akhmerov09,Seradjeh09,Garate10,Yu10}.

Here we confirm unequivocally that the weak antilocalization effect can be used to differentiate the surface transport from transport dominated by bulk carriers. This is demonstrated on Bi$_2$Se$_3$ thin films with carrier densities that can be tuned over a wide range with a back-gate. When the transport is not in the topological regime, the magnetoconductivity can be described by a single component Hikami-Larkin-Nagaoka (HLN) equation~\cite{Hikami80}. This description is found to be valid for a remarkably wide range of electron densities (0.8-8.6$\times10^{13}$cm$^{-2}$) in samples with the Fermi energy located inside the conduction band, even if electron-electron interactions are taken into account. In contrast, in a regime where the electronic system is split up into an electron layer at the top surface and a hole layer at the bottom, the magnetoconductivity deviates strongly from the single-component HLN equation. Our analysis provides a convenient method for detecting decoupled surface transport. It complements existing techniques based on quantum oscillations that are limited to samples of high carrier mobilities~\cite{QuantumOsc} or samples with a quasi-1D geometry~\cite{Peng10}.

The Bi$_2$Se$_3$ thin films were grown on SrTiO$_3$(111) substrates with molecular beam epitaxy~\cite{ZhangG11}.
The dielectric properties of ${\rm SrTiO}_{3}$ are well suited for gating purposes and
 the carrier density in these devices can be varied by at least $2\times10^{13}$\,cm$^{-2}$~\cite{ChenJ10}.  All of the samples used in this work were patterned into 50\,$\mu$m wide Hall bars with photolithography, followed by Ar plasma etching (Fig.\,1 inset). This eliminates uncertainties in evaluating resistivities encountered in previous transport studies due to the influence of electrical contacts or the irregular shape of the sample. A set of more than ten samples with thicknesses between 5 and 20\,nm has been measured. Most of the samples have a back-gate deposited at the bottom of the substrate, and a few of them are equipped in addition with a top-gate. The latter was deposited on an AlO$_x$ layer prepared with atomic layer deposition. Transport measurements were carried out in cryostats with temperatures as low as 10\,mK and magnetic fields up to 18\,T.

\begin{figure}
\centering
\includegraphics*[width=7.5 cm]{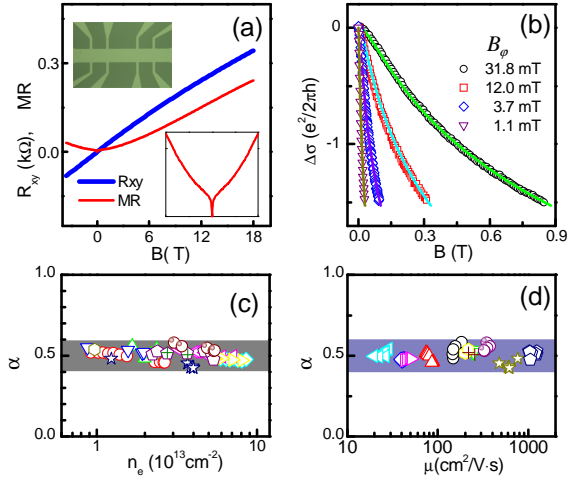}
\caption{(color online) (a) Magnetoresistance (MR, defined as $\rho_\mathrm{xx}(B)/\rho_\mathrm{xx}(0)-1$) and Hall resistance of a typical Bi$_2$Se$_3$ thin film at $T=1.2$\,K. The low field MR (between $\pm3$\,T) is shown more clearly in the lower inset. The upper inset is an optical image of the Hall bar used for the transport measurements. (b) Magnetoconductivity data (symbols) from four samples fitted with Eq.\,(1) (lines). The dephasing field $B_\phi$ varies by nearly a factor of 30. Extracted values of $\alpha$ for 10 samples are plotted as a function of electron density $n_e$ and mobility $\mu$ in (c) and (d), respectively. Both are evaluated based on the low field transport measurements.} \label{Fig1}
\end{figure}

Fig.\,1 displays typical magnetotransport data.
All of the samples show a positive magneto-resistance with a sharp cusp around zero magnetic field, consistent with previous measurements of Bi$_2$Se$_3$ thin films~\cite{ChenJ10,Checkelsky10,HeT10,LiuM10,WangJ10}. As demonstrated in Fig.\,1(b), the low field magnetoconductivity, defined as $\Delta\sigma(B)=\sigma_\mathrm{xx}(B)-\sigma_\mathrm{xx}(0)$, can be fitted well with the HLN equation in the strong SOC limit, i.e. when $\tau_\phi\gg \tau_\mathrm{so},\tau_e$:
\begin{equation}
\label{eq:WALtopo}
\Delta\sigma(B)\simeq
-\alpha\cdot\frac{e^2}{\pi h}
\left[
\psi\left(\frac{1}{2}+\frac{B_\phi}{B}\right)
-\ln\left(\frac{B_\phi}{B}\right)
\right],
\end{equation}
where $\tau_\mathrm{so}$ ($\tau_e$) is the spin-orbit (elastic) scattering time, $\psi$ is the digamma function, $B_\phi=\hbar/(4De\tau_\phi)$ is a characteristic field related to the dephasing time $\tau_\phi$, $D$ is the diffusion constant and $h$ is the Planck constant. The coefficient $\alpha$ takes a value of 1/2 for a traditional 2D electron system with strong spin-orbit coupling. The same value is expected for the electron transport on one surface of a 3D TI with a single Dirac-cone~\cite{ChenJ10}.

Fig.\,1(c) shows that the extracted $\alpha$ values are distributed in a narrow range near 1/2 for 10 samples with 2D electron densities $n_e$ spreading from 0.8 to 8.6$\times10^{13}$cm$^{-2}$~\cite{YangF11}. No correlation is found between $\alpha$ and the electron mobility $\mu$, which varies nearly two orders of magnitude (Fig.\,1(d)). Based on angle-resolved photoemission measurements~\cite{Hsieh09b}, the top and bottom surfaces of a Bi$_2$Se$_3$ thin film can only accommodate a total electron density of $\sim0.5\times10^{13}$\,cm$^{-2}$ even if the Fermi energy reaches the bottom of the conduction band. Thus we anticipate a significant number of bulk electrons (or quasi-2D electrons with parabolic dispersion) for the above range of $n_e$. The nonlinear Hall resistivity curves (see e.g.\,Fig.\,1(a)) also suggest the coexistence of multiple charge carrier types.
Even if so, the analysis of the weak antilocalization effect itself at small magnetic fields yields values of $\alpha$ close to 1/2. In this magnetic field regime where the antilocalization effect is observed, these samples do behave like 2D systems with a single type of charge carrier. This can only be understood when there is a strong mixing between the surface and the bulk electron states or when the dephasing field of one of the conducting components (i.e.\ the bulk or top/bottom surface) is much smaller than those of the others. We note that it was demonstrated long ago that the two-valley 2D electron system confined in a Si inversion layer displays $\alpha$ values close to that for a single-valley system and not the one expected for two independent valleys~\cite{Kawaguchi82}. Fukuyama attributed it to intervalley scattering~\cite{Fukuyama85}. Similar physics might take place here because of considerable scattering between the surface and bulk states when the Fermi energy is located in the conduction band.

The robustness of $\alpha\simeq1/2$ is at first sight surprising, because sources other than weak antilocalization may also contribute to the low field magnetoconductivity. In the non-interacting electron picture, the Zeeman energy, which was not considered in deriving the HLN equation, is known for mixing the spin singlet and triplet states and hence  suppressing the weak antilocalization effect~\cite{Maekawa81}. The corresponding correction to $\Delta\sigma(B)$ is determined by the ratio $\gamma=E_Z/E_\mathrm{so}=g\mu_B B/(\hbar\tau_\mathrm{so}^{-1})$, where $g$ is the electron $g$-factor. The Zeeman energy also causes an extra change in $\Delta\sigma(B)$ if the electron-electron interaction is not negligible in the diffusion channel~\cite{Lee85,Altshuler82}. The corresponding correction to the conductivity is
$\Delta\sigma_I(B)=\frac{e^2}{\pi h}\frac{\widetilde{F}^\sigma}{2}g_2(\tilde{h})$ with $\tilde{h}=E_Z/k_B T$, where $\widetilde{F}^\sigma$ is a parameter reflecting the strength of dynamically screened Coulomb interaction.


\begin{figure}
\centering
\includegraphics*[width=8 cm]{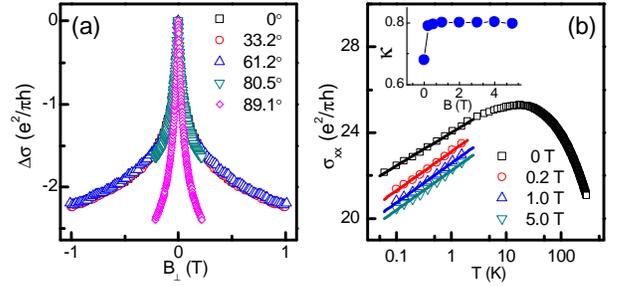}
\caption{(color online) (a) Magnetoconductivity $\Delta\sigma(B)$ at $T=2$\,K and $V_G=$ plotted as a function of perpendicular magnetic field $B_\perp$ for several tilt angles. $\theta=0$ refers to $\mathbf{B}$ perpendicular to the thin film plane.  (b) Temperature dependencies of $\sigma_\mathrm{xx}$ (open symbols) recorded at $B=$0, 0.2, 1, and 5\,T. The straight lines are linear fits of $\sigma_\mathrm{xx}$ to $\ln T$. In the upper inset, the slope, defined as $\kappa=(\pi h/e^2)d\sigma_\mathrm{xx}(B,T)/d(\ln T)$, is plotted as a function $B$. The electron density is about $2\times10^{13}$cm$^{-2}$ so that $E_F$ is located in the conduction band.} \label{Fig2}
\end{figure}

 Since the electron $g$-factor of bulk Bi$_2$Se$_3$ is quite large~\cite{Koehler75}, one would expect sizable Zeeman corrections to $\Delta\sigma(B)$. This appears to be in contradiction with the data recorded in tilted magnetic fields and plotted in Fig.\,2(a). The low field magnetoconductivity exhibits very little angular dependence for tilt angles less than 80\,$^\circ$. Considering that $E_Z$ nearly doubles (triples) for $\theta=$60$^\circ$ (70$^\circ$) with respect to the zero-tilt case, we conclude that the influence of the Zeeman energy can be neglected in case of zero- or small tilts. In the non-interacting regime, this can be understood as a consequence of strong SOC, and hence small $\gamma$ for the fields of interest. Also in the regime where e-e interactions are important, the strong SOC suppresses the Zeeman contribution. The Zeeman term was derived under the assumption  of weak SOC~\cite{Lee85}. Theories~\cite{Altshuler82,Lee85,Altshuler85} and experiments~\cite{Sahnoune92} on other materials have clearly shown that strong SOC can diminish and even entirely suppress the Zeeman-split term in the diffusion channel.

The effects of strong SOC are further manifested in the temperature dependence of $\sigma_\mathrm{xx}$ displayed in Fig.\,2(b). The slope of the $\Delta \sigma(B)$-$\ln T$ plot, defined as $\kappa=(\pi h/e^2)d\sigma_\mathrm{xx}(B,T)/d(\ln T)$, is nearly constant for $B$=0.2-5\,T. Both weak antilocalization and  e-e interaction can cause the $\ln T$ dependence~\cite{Bergmann84,Lee85}. The weak antilocalization effect however only produces a pronounced $T$-dependence to $\sigma_\mathrm{xx}$ at zero or low magnetic fields. The nearly constant slope at $B>0.2$\,T can be attributed to the strong SOC, which suppresses the triplet terms~\cite{Altshuler85,Ostrovsky10}. They would otherwise produce $\ln T$ corrections proportional to $\widetilde{F}^\sigma$~\cite{Lee85,Altshuler85}. Hence, $\kappa=1$ is expected~\cite{Ostrovsky10} for sufficiently large $B$. The observed $\kappa\simeq0.8$ is slightly smaller. This deviation may originate from other sources such as the corrections in the Cooperon channel~\cite{MirlinNote}. Nevertheless, the nearly constant $\kappa$ for $B=0.2$-$5$\,T indicates that the e-e interaction does not induce significant corrections to $\Delta \sigma(B)$ in lower fields, where the weak antilocalization is pronounced. As to the zero-field conductivity, the combined effects of e-e interactions and weak antilocalization lead to a $\ln T$ dependence with $\kappa=1-1/2$=1/2, which is qualitatively in agreement with our data and others~\cite{LiuM10,WangJ10}. Taken together with the tilted-field data, we can conclude that the single-particle HLN equation in the strong SOC limit (Eq.\,(1)) can provide a good description of the low field magnetoconductivity for a conducting 2D channel with strong SOC or one surface of a 3D TI.

Now we are in a position to use the low field magneto-transport as a tool to analyze the influence of a negative gate voltage.  Fig.\,3(a) shows Hall data from one of the samples with large density tunability. It is a 10\,nm thick undoped Bi$_2$Se$_3$ film. The electron density at $V_G$=0, estimated from the low field Hall resistance, is about $2.7\times10^{13}$\,cm$^{-2}$. The Hall resistance, $R_\mathrm{xy}(B)$, increases as $V_G$ decreases. It reaches a maximum at $V_G=-125$\,V, which would correspond to an electron density of $n_e\approx0.3\times10^{13}$\,cm$^{-2}$. Further decrease in $V_G$ leads to smaller $R_\mathrm{xy}(B)$ and even reversal of its sign. For $V_G<-150$\,V, the Hall curves become strongly nonlinear. The high field Hall coefficient is plotted in Fig.\,3(b) and depends non-monotonously on gate voltage.
Also shown is the longitudinal resistivity at $B=0$, denoted as $\rho_\mathrm{xx}(0)$ throughout this paper. It also exhibits a non-monotonic dependence on $V_G$. This, together with the fact that the Hall coefficient $R_H$ does not reach a minimum, points to the coexistence of electrons and holes for large negative gate voltages. It is noteworthy that the maximum in $\rho_\mathrm{xx}(0)$ appears at a $V_G$ smaller than that of the $R_H$  maximum. Therefore, the crossover from the pure electron system to the electron-hole system must take place before the appearance of the $\rho_\mathrm{xx}(0)$ maximum.

\begin{figure}
\centering
\includegraphics*[width=7.5 cm]{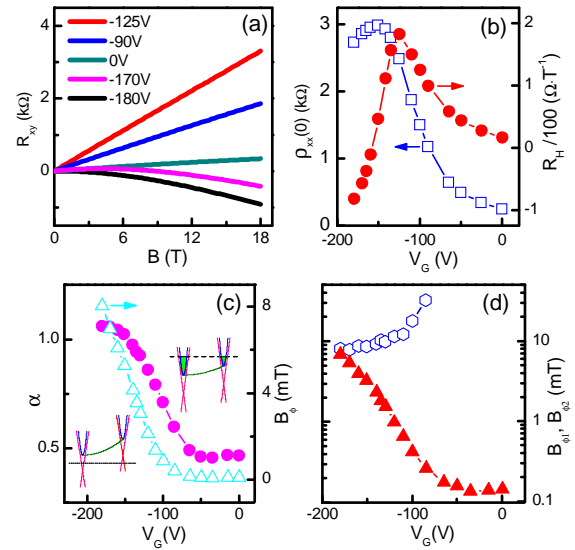}
\caption{(color online)
(a) Hall resistance curves for $V_G=-125, -90, 0, -170, -180$\,V (from top to bottom).  (b) Gate-voltage dependence of $\rho_\mathrm{xx}$ at $B=0$ and high field Hall coefficient, defined as $dR_\mathrm{xy}/dB$ and fitted from the data in $B=16$-18\,T. (c) Gate-voltage dependence of $\alpha$ and $B_\phi$ obtained from fits to Eq.\,(1). Shown in the left and right insets are the band diagrams for large and small negative gate voltages, respectively. The top (bottom) surface is depicted on the left(right). (d) $V_G$ dependence of $B_{\phi1}$ (hexagons) and $B_{\phi2}$ (triangles) obtained from fits to Eq.\,(2). $B_{\phi1} $ and $B_{\phi2}$ can be assigned to the bottom and the top surfaces, respectively. This 10\,nm thick sample only has a back-gate.
}
\label{Fig3}
\end{figure}

 For the gate voltages smaller than that at the $R_\mathrm{xy}$ maximum, the Fermi energy on the bottom and top surfaces are expected to lie below and above the Dirac point, respectively, even though the precise position of $E_F$ is not known. As a consequence, the Fermi energy in the bulk (or at least part of the bulk) must be located in the band gap. The nearly one order of magnitude increase in $\rho_\mathrm{xx}(0)$ as $V_G$ is lowered from 0 to $-150$\,V is much larger than what has been reported for cleaved Bi$_2$Se$_3$ flakes cleaved on SiO$_{2}$/Si substrates\cite{Checkelsky10}.
 The significantly enhanced $\rho_\mathrm{xx}(0)$ is an encouraging signature that much of the bulk conductivity can be suppressed. It can reach values as high as $\sim h/e^2$~\cite{ChenJ10}.

The magnetoconductivity also exhibits a strong  gate-voltage dependence, especially for $V_G<-50$\,V. Best fits to Eq.\,(1) yield the data plotted in Fig.\,3(c). The most striking feature is that $\alpha$ is close to 1/2 for $V_G>-70$\,V, and it increases to values close to 1 for $V_G<-140$\,V. In the crossover region ($-70$ to $-140$\,V), the Bi$_2$Se$_3$ thin film undergoes a transition from a low density electron system to a separated electron-hole system. Hence, for large negative gate voltages, a fit of the magnetoconductivity data to a two-component HLN equation is more appropriate:
\begin{equation}
\label{eq:WAL2}
\Delta\sigma(B)\simeq
-\frac{e^2}{2\pi h}\sum_{i=1}^{2}
\left[
\psi\left(\frac{1}{2}+\frac{B_{\phi i}}{B}\right)
-\ln\left(\frac{B_{\phi i}}{B}\right)
\right]
\end{equation}
Here $B_{\phi1}$ and $B_{\phi2}$ are dephasing fields for conducting components 1 and 2, respectively. As shown in Fig.\,3(d), they have opposite dependencies on $V_G$. $B_{\phi1}$($B_{\phi2}$) decreases(increases) as $V_G$ is lowered. They approach approximately the same value for large negative gate voltages. The dephasing field is proportional to $(D\tau_\phi)^{-1}\propto (v_F^2\tau_e\tau_\phi)^{-1}$, so $B_\phi$ is expected to increase for the electron component on the top surface, while it should  decrease for the holes at the bottom surface (interface) with decreasing  $V_G$. Therefore, the two curves with larger and smaller values of $B_\phi$ in Fig.\,3(d) could be assigned to the bottom and top surfaces, respectively.

The observation that $\alpha$ increases toward 1 based on fits of the magnetoconductivity data to the single-component HLN equation (Eq.\,1) implies that the top and bottom surfaces of the film make separate contributions to the conductivity~\cite{note_on_Ong}. Obtaining $\alpha$ values close to 1, however, not only requires two decoupled conduction channels, but also demands that both conduction channels have nearly identical dephasing fields. This is in general hard to achieve, in particular for samples where the gate tunability is not sufficient or the substrate surface is too rough~\cite{ChenJ10}. Caution should also be taken to ensure that the transport is in the diffusive and weakly disordered ($k_F l\gg 1$) regime for which the HLN equation is valid. For highly resistive samples, e.g.\ $\rho_\mathrm{xx} \sim h/e^2$ as shown in Ref.~\cite{ChenJ10}, the condition $k_Fl\gg 1$ is no longer satisfied.

In conclusion, it is possible to identify the surface transport of 3D topological insulators from magnetoconductivity measurements when transport takes place in the weakly disordered, diffusive regime. The use of a high-$k$ dielectric such as SrTiO$_3$ for back-gating enables us to tune the transport properties of both the top and the bottom surfaces. This device geometry is particularly useful for the future exploration of hybrid devices in which a topological insulator is interfaced with a superconductor~\cite{Fu09,Akhmerov09} or a  ferromagnet~\cite{Fu09,Akhmerov09,Garate10}).

We are grateful to L. Fu, D. Goldhaber-Gordon, I.~V. Gornyi, P.~M. Ostrovsky, V. Sacksteder, K. von Klitzing, X.~C. Xie, P. Xiong, and in particular A.~D. Mirlin for valuable discussions. The work was supported by MOST-China, NSF-China, and Chinese Academy of Sciences and German Ministry of Science \& Education.


\begin{thebibliography}{}
\bibitem{note_on_Ong}Qualitatively similar variation of $\alpha$ with $V_G$ was also reported in Ref.\,[8], in which the weak antilocalization signal is however superimposed on fluctuating background of unknown origin. The non-Hall-bar device-geometry also precluded accurate determination of $\alpha$ in that work.
\end{thebibliography}

\end{document}